\documentclass[reprint,pra,groupedaddress,superscriptaddress,nofootinbib,aps,longbibliography,amsmath,amssymb,10pt]{revtex4-2}
\usepackage{url}
\usepackage{graphicx}
\usepackage{placeins}
\usepackage{dcolumn}
\usepackage{bm}

\usepackage{amsmath}
\usepackage{physics}
\usepackage{dsfont}
\usepackage{tikz,pgfplots}
\usepackage{amsthm}
\usepackage{enumerate}
\usepackage{subfigure}
\usepackage{bbold}
\usepackage{makecell}
\usepackage{array}
\usepackage{tikz}
\usetikzlibrary{calc}

\newcommand{\nxrightarrow}[1]{%
  \mathrel{%
    \begin{tikzpicture}[baseline=-0.7ex]
      \node[inner sep=0.7pt] (arrow) {$\xrightarrow{#1}$};
      \draw[line width=0.35pt] 
        ($(arrow.center)-(2pt,6pt)$) -- ($(arrow.center)+(1pt,-1pt)$);
    \end{tikzpicture}%
  }%
}

\newcommand{\nlrightarrow}[1]{%
  \mathrel{%
    \begin{tikzpicture}[baseline=-0.7ex]
      \node[inner sep=0.7pt] (arrow) {$\xleftarrow{#1}$};
      \draw[line width=0.35pt] 
        ($(arrow.center)-(2pt,6pt)$) -- ($(arrow.center)+(1pt,-1pt)$);
    \end{tikzpicture}%
  }%
}

\makeatletter
\newcommand{\thickhline}{%
    \noalign {\ifnum 0=`}\fi \hrule height 1pt
    \futurelet \reserved@a \@xhline
}
\newcolumntype{"}{@{\hskip\tabcolsep\vrule width 1pt\hskip\tabcolsep}}
\makeatother
\usepackage{multirow}
\usepackage{tabu}
\usepackage{textcomp,booktabs}
\usepackage{colortbl}
\usepackage[T1]{fontenc}

\newcommand{\mO}{\mathcal{O}}

\makeatletter
\newcommand*{\rom}[1]{\expandafter\@slowromancap\romannumeral #1@}
\makeatother

\makeatletter
\newcommand{\xleftrightarrow}[2][]{\ext@arrow 3359\leftrightarrowfill@{#1}{#2}}
\makeatother

\makeatletter
\newcommand{\nxleftrightarrow}[2][]{\ext@arrow 3359\nleftrightarrowfill@{#1}{#2}}
\makeatother

\usepackage{amsthm}
\newtheorem*{thm*}{Theorem}
\newtheorem{thm}{Theorem}[section]
\newtheorem{lem}[thm]{Lemma}
\newtheorem{cor}[thm]{Corollary}
\newtheorem{definition}{Definition}[section]

\usepackage{hyperref}
\pgfplotsset{compat=1.17}
\begin{document}


\title{
Reusability of Quantum Catalysts
}

\author{Haitao Ma}
\affiliation{School of Mathematical Sciences, Harbin Engineering University, Nantong Street, Harbin 150001, China}

\author{Yantong Li}
\affiliation{School of Mathematical Sciences, Harbin Engineering University, Nantong Street, Harbin 150001, China}

\author{Yingchun Kang}
\affiliation{School of Mathematical Sciences, Harbin Engineering University, Nantong Street, Harbin 150001, China}

\author{Bing Yu}
\affiliation{School of Mathematics and Systems Science, Guangdong Polytechnic Normal University, Guangzhou 510665, China}

\author{Junjing Xing}
\email{xingjunjing2017@hrbeu.edu.cn}
\affiliation{School of Mathematical Sciences, Harbin Engineering University, Nantong Street, Harbin 150001, China}

\author{Zhaobing Fan}
\email{fanzhaobing@hrbeu.edu.cn}
\affiliation{School of Mathematical Sciences, Harbin Engineering University, Nantong Street, Harbin 150001, China}

\author{Yunlong Xiao}
\email{mathxiao123@gmail.com}
\affiliation{Institute of High Performance Computing (IHPC), Agency for Science, Technology and Research (A*STAR), 1 Fusionopolis Way, \#16-16 Connexis, Singapore 138632, Republic of Singapore}
\affiliation{Quantum Innovation Centre (Q.InC), Agency for Science Technology and Research (A*STAR), 2 Fusionopolis Way, Innovis \#08-03, Singapore 138634, Republic of Singapore}

\date{\today}
             
\begin{abstract} 
Quantum catalysts enable transformations that otherwise would  be forbidden, offering a pathway to surpass conventional limits in quantum information processing. 
Among them, embezzling catalysts stand out for achieving near-perfect performance while tolerating only minimal disturbance, bridging the gap between ideal and practical catalysis. 
Yet, this superior capability comes at a cost: Each use slightly degrades the catalyst, leading to an inevitable accumulation of imperfection. 
This gradual decay defines their most distinctive property -- reusability -- which, despite its fundamental importance, remains largely unexplored. 
Here, we establish a quantitative framework to characterize the operational lifetime of embezzling catalysts, focusing on their role in entanglement distillation and extending the analysis to quantum teleportation. 
We show that the catalytic advantage inevitably diminishes with repeated use, deriving bounds on the maximum effective reuse rounds for a desired performance gain. 
Our results uncover the finite reusability of catalysts in quantum processes and point toward sustainable strategies for quantum communication.
\end{abstract}

\maketitle


\section{Introduction}

Catalysis lies at the heart of transformation in chemistry, biology, and materials science, enabling reactions and processes that otherwise would be unattainable or prohibitively slow~\cite{CHEN2024147853,https://doi.org/10.1002/aenm.202304099,https://doi.org/10.1002/advs.202306979}. 
Extending this idea to the quantum domain, quantum catalysts~\cite{PhysRevLett.132.140402,PhysRevLett.132.180202,PhysRevLett.132.260403,PhysRevLett.133.050601,PhysRevLett.133.140201} act not on chemical species but on quantum information processing itself, allowing state transformations and communication protocols that exceed what is achievable through standard quantum operations alone.
Once a theoretical curiosity, catalytic behavior now surfaces across entanglement theory~\cite{Datta2024entanglement,Zanoni2024complete,PhysRevLett.133.250201}, quantum thermodynamics~\cite{PhysRevX.11.011061,PhysRevLett.132.200201,PhysRevLett.134.160402}, quantum communication~\cite{PhysRevLett.127.080502,Xing2024Teleportation,Li_2025}, and even Bell nonlocality~\cite{5dth-7zm8} -- showing that with the subtle aid of a quantum catalyst, the limits of conventional quantum processes can be genuinely surpassed.

Depending on their structure, quantum catalysts can be classified into distinct categories. 
The most ideal form is the exact catalyst~\cite{PhysRevLett.83.3566, PhysRevA.64.042314, PhysRevA.71.042319,Turgut_2007, Marvian_2013, doi:10.1038/ncomms7383, PhysRevA.93.042326}, which assists a quantum process, yet remains completely separable from the system of interest and returns precisely to its original state once the operation is complete. 
Such catalysts enhance the performance of quantum information processing without being consumed, allowing in principle unlimited reuse.
However, like in chemistry, perfection is elusive. 
For a given transformation, for example, entanglement distillation or quantum communication, there is often no general method to design a suitable catalyst, especially when its dimension is finite and the desired improvement is fixed.

The parallel with classical catalysis goes deeper. 
Chemical catalysts~\cite{FORZATTI1999165}, though regenerable through heat, oxidation-reduction cycles, or purification, inevitably lose some activity with each round, leading to a finite operational lifetime. 
Inspired by this reality, the notion of an embezzlement catalyst arose, a quantum system that tolerates a small disturbance to itself after catalysis~\cite{PhysRevA.67.060302,PhysRevLett.111.250404,PhysRevLett.113.150402,PhysRevA.90.042331,doi:10.1073/pnas.1411728112,doi:10.1007/s00037-015-0098-3,Ng_2015,7377103,10.1063/1.4974818,PhysRevLett.118.080503, PhysRevLett.121.190504,PhysRevA.100.042323,doi:10.1038/s41534-022-00608-1,PhysRevX.13.011016,10086536,10121557,Luijk2023covariantcatalysis,PRXQuantum.4.040330,vanluijk2024embezzlement,PhysRevLett.133.261602}. 
This concept bridges ideal and practical catalysis, capturing the delicate balance between performance and degradation that defines catalytic behavior in the quantum realm. 
Remarkably, such quantum catalysts can drive quantum teleportation that verge on perfection while perturbing themselves by only an infinitesimal amount~\cite{Xing2024Teleportation}.

Previous investigations have mainly examined embezzling catalysts in the single-use setting, where the catalyst participates in only one round of transformation before being reset or discarded. 
Such a framework makes it difficult to distinguish the genuine catalytic effect from the simple addition of extra resources. 
What truly defines a catalyst, however, is its reusability -- its ability to drive successive transformations while largely preserving its functional capacity. 
Despite its importance, this fundamental aspect has remained largely unexplored. 
In this work, we take entanglement distillation, a central process in quantum communication and computation, as a model task to systematically investigate the operational power and limitations of embezzling catalysts. 
We address a key question: given a measurable performance improvement, how many rounds can the catalyst be reused before its advantage disappears? 
Our results reveal that the benefit provided by an embezzling catalyst inevitably diminishes with repeated use, defining a finite catalytic lifetime that quantifies the trade-off between enhancement and degradation. 
Extending our analysis to quantum teleportation, we demonstrate how understanding catalytic reusability can inform the design of more practical and resource-efficient quantum communication protocols.

The structure of this work is as follows. Sec.~\ref{sec:catalyst} introduces the key concepts underlying quantum catalysis and outlines the operational framework used in subsequent analyses. 
Sec.~\ref{sec:distillation} then presents a detailed study of the reusability of two classes of embezzling catalysts in entanglement distillation, deriving bounds on the maximum number of effective reuse rounds achievable under a prescribed fidelity improvement threshold. 
Building on these results, Sec.~\ref{sec:teleportation} extends the framework to quantum teleportation, establishing reusable catalytic protocols that highlight the practical relevance of the theory. 
Finally, Sec.~\ref{sec:conclusion} summarizes the main findings and discusses broader implications for catalytic quantum communication.


\section{Quantum catalysts}\label{sec:catalyst}
Just as catalysts in chemical reactions or enzymes in biochemical processes enable transformations that would otherwise be unattainable, quantum catalysts open new pathways for state conversion under otherwise constrained operational settings. 
In this section, we introduce the basic ideas of quantum catalysis in entanglement theory, highlighting the notion of embezzling catalysts -- a generalized form of catalysis that tolerates a slight alteration after use. 
This subtle relaxation, far from being a drawback, proves essential for realistic implementations of quantum information processing. 
A broader perspective on the development of quantum catalytic frameworks can be found in Ref.~\cite{RevModPhys.96.025005}.

Let $\rho$ and $\sigma$ be two bipartite quantum states that are mutually inconvertible under local operations and classical communication (LOCC), i.e., neither can be transformed into the other within entanglement theory
\begin{align}\label{eq:ent-no-locc}
    \rho
    \nxrightarrow{\text{LOCC}}
    \sigma
    \quad
    \text{and}
    \quad
    \rho
    \nlrightarrow{\text{LOCC}}
    \sigma.
\end{align}
Remarkably, the introduction of an auxiliary catalytic state $\tau$ can circumvent this limitation, enabling the transformation
\begin{align}\label{eq:ent-cata-locc}
    \rho\otimes\tau
    \xrightarrow{\text{LOCC}}
    \sigma\otimes\tau,
\end{align}
while leaving $\tau$ exactly unchanged. Such a process is referred to as {\it exact catalysis}~\cite{PhysRevLett.83.3566, PhysRevA.64.042314, PhysRevA.71.042319,Turgut_2007, Marvian_2013, doi:10.1038/ncomms7383, PhysRevA.93.042326}.

In many entanglement-based quantum information protocols, such as quantum teleportation and quantum repeaters, the requirement of exact catalyst recovery imposes stringent constraints on both the attainable performance and the advantages that catalysis can provide. 
{\it Embezzling catalysis}~\cite{PhysRevA.67.060302,PhysRevLett.111.250404,PhysRevLett.113.150402,PhysRevA.90.042331,doi:10.1073/pnas.1411728112,doi:10.1007/s00037-015-0098-3,Ng_2015,7377103,10.1063/1.4974818,PhysRevLett.118.080503, PhysRevLett.121.190504,PhysRevA.100.042323,doi:10.1038/s41534-022-00608-1,PhysRevX.13.011016,10086536,10121557,Luijk2023covariantcatalysis,PRXQuantum.4.040330,vanluijk2024embezzlement,PhysRevLett.133.261602}, which tolerates a slight deviation in the catalyst state, offers a more practical and versatile alternative, enabling a substantial enhancement in achievable performance. 
We now introduce a formal definition of this central notion.

\begin{definition}
[{\bf Embezzling Catalysts~\cite{Datta_2023}}]
\label{def:AC}
A state transition from $\rho_A$ to $\sigma_A$ on system $A$ is said to be embezzling catalytic with respect to a set of free operations $\mO$ if there exists a catalytic system $C$ prepared in a state $\tau_C$ and a free operation $\Lambda \in \mO$ such that
\begin{align}
    D\left(\Lambda(\rho_A\otimes \tau_C), \sigma_A\otimes\tau_C\right)
    \leqslant
    \varepsilon,
\end{align}
and
\begin{align}\label{eq:D-catalyst}
    D\left(\Tr_A\left[\Lambda(\rho_A\otimes \tau_C)\right], \tau_C\right)
    \leqslant
    \delta,
\end{align}
where $\varepsilon, \delta > 0$ are error parameters and $D$ denotes a suitable distance measure, such as the trace distance~\cite{PhysRevLett.122.141602,Chen2021} or the purified distance~\cite{rastegin2006sine,tomamichel2013framework}. The state $\tau_C$ is called an embezzling catalyst, as it facilitates the transformation from $\rho_A$ to $\sigma_A$ while remaining only slightly altered -- its deviation controlled by the small error parameters $\delta$.
\end{definition}


\section{Catalyst Reuse}\label{sec:distillation}
Entanglement underpins the unparalleled performance of quantum communication compared with its classical counterpart. 
In realistic settings, however, this advantage is inevitably degraded by environmental noise, motivating the use of entanglement distillation to recover high-quality entangled states. 
Incorporating embezzling catalysts can further enhance the efficiency of such protocols and has found broad relevance across quantum communication tasks. 
Yet, a key aspect of catalysis -- its reusability -- remains largely unexplored. 
In this section, we address this question and establish quantitative limits on how many times an embezzling catalyst can be reused under fixed accuracy constraints, thereby revealing the fundamental bounds of catalytic resource recycling in quantum information processing.


\begin{figure*}
    \centering
    \includegraphics[width=1\textwidth]{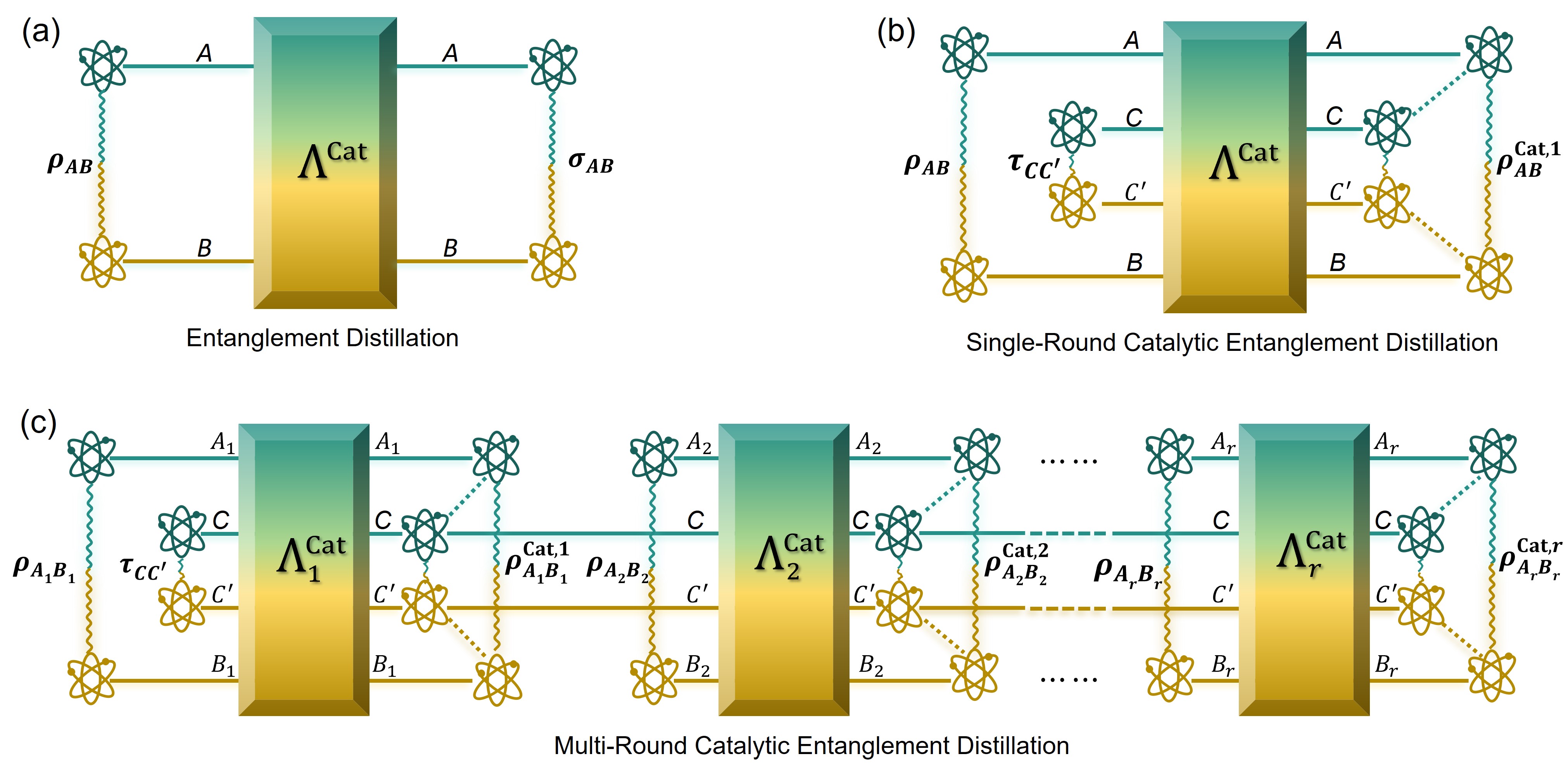}
    \caption{{\bf Schematic of Catalytic Entanglement Distillation.}
    (a) shows the standard protocol without auxiliary entangled resources, where a single noisy state $\rho_{AB}$ is purified solely through a LOCC operation $\Lambda$. 
    (b) depicts a single-round catalytic distillation protocol assisted by an embezzling state $\tau_{CC'}$, which provides auxiliary entanglement while remaining nearly unchanged. 
    (c) extends this to a multi-round catalytic distillation scheme, in which the catalyst is reused across successive rounds. For each round $i = 1, 2, \dots, r$, the operation $\Lambda^{\text{Cat}}_i$ is identical to that in (b) and acts on the subsystem $A_i C C' B_i$, with each round starting from an identical input state $\rho_{AB}$.
    }
    \label{fig:Entdistillation}
\end{figure*}

\subsection{Catalytic Entanglement Distillation}
Entanglement distillation plays a central role in quantum communication, serving as the essential mechanism for recovering high-quality entanglement that has been degraded by noise~\cite{Kwiat2001,Devetak2005,Dong2008,Takahashi2010,Kalb2017,Leditzky2018,PhysRevLett.127.040506,Lami2024,PhysRevLett.132.180201}. 
By applying LOCC operations, multiple copies of weakly entangled states can be converted into a smaller number of maximally entangled pairs. 
This process underlies the reliable operation of quantum teleportation, repeaters, and networks. 
Of particular importance is single-shot entanglement distillation~\cite{Brandao2011,Regula2019,Fang2019,ArnonFriedman2019,PhysRevLett.127.150503,glc7-xy8t}, which extracts high-quality entanglement from a single copy of a noisy resource rather than relying on asymptotic many-copy limits (see Fig.~\ref{fig:Entdistillation}(a)). 
Such a setting is especially relevant in practical scenarios where resource states are limited or repeated preparation is infeasible, enabling immediate and efficient use of entanglement in quantum communications. 
Moreover, this single-copy framework offers a natural arena to explore catalytic and embezzling phenomena that fundamentally reshape our understanding of distillation efficiency, resource conversion, and reusability.

Given a bipartite quantum state $\rho_{AB}$ acting on systems $A$ and $B$, its entanglement can be characterized by the entanglement fidelity,
\begin{align}\label{eq:Entfidelity}
    F(\rho_{AB}):= \Tr[\rho_{AB} \cdot \phi^+_{d,AB}],
\end{align}
which quantifies how closely state $\rho_{AB}$ approximates a $d$-dimensional maximally entangled state $\phi^+_{d}$.
To assess the single-shot distillable entanglement, we consider the optimal transformation of $\rho_{AB}$ under LOCC operations. 
In this setting, one seeks the highest attainable fidelity with $\phi^+_{d}$, optimized over all LOCC protocols. 
The resulting quantity captures the maximum entanglement that can be distilled from a single copy of $\rho_{AB}$ and is formally expressed as
\begin{align}\label{eq:Fmax}
    F_{\max}(\rho_{AB}):= 
    \max_{\Lambda\in \text{LOCC(A:B)}} \Tr[\Lambda(\rho_{AB}) \cdot \phi^+_{d,AB}].
\end{align}

In a recent work~\cite{Xing2024Teleportation}, the authors demonstrated that the fundamental limit in Eq.~\eqref{eq:Fmax} can be surpassed once an embezzling catalyst is introduced into the distillation process. 
The catalyst effectively supplies a hidden reservoir of entanglement that can be gently borrowed and almost perfectly returned, enabling transformations that would otherwise be forbidden under LOCC. 
Remarkably, even a finite-dimensional catalyst was found sufficient to distill nearly perfect maximally entangled pairs, pushing the achievable entanglement fidelity arbitrarily close to unity. 
To uncover this phenomenon, the authors developed two distinct constructions of embezzling catalysts -- one grounded in the convex-split lemma and the other inspired by the canonical embezzling state. 
For clarity, we briefly review these two approaches in the following lemmas, beginning with the convex-split–lemma–assisted (CSLA) distillation protocol.

\begin{lem}
[{\bf CSLA Distillation~\cite{Xing2024Teleportation}}]
\label{lem:cs}
Given a bipartite quantum state $\rho_{AB}$ of local dimension $d$ and a threshold $\varepsilon>0$, we select a positive bipartite state $\tau_{AB}$ satisfying $F(\tau)\geqslant 1- \varepsilon/4$.
Let
\begin{align}
    k:= D_{\rm{max}}(\rho\parallel\tau)
\end{align}
denote the max-relative entropy of $\rho_{AB}$ with respect to $\tau_{AB}$.
Under these conditions, one can construct a catalytic state
\begin{align}\label{eq:CSLA-Catalyst}
    \tau^{CS}:= \tau^{\otimes n-1},
\end{align}
with
\begin{align}
    n:= \left\lceil \frac{2^{k+2}}{\varepsilon} \right\rceil,
\end{align}
which serves as a convex-split catalyst, enabling the desired transformation
\begin{align}\label{CLA}
    F\left(\Tr_{CC^{'}}[\Lambda^{CS}(\rho\otimes \tau^{CS})] \right)\geqslant 1-\varepsilon.
\end{align}
The corresponding LOCC operation $\Lambda^{CS}$ in Eq.~\eqref{CLA} is defined as
\begin{align}\label{eq:lmdCS}
\Lambda^{CS}(\rho\otimes\tau^{\otimes n-1}):=
\frac{1}{n}\sum_{t=1}^{n}\tau_{n}\otimes\cdots\otimes\rho_{t}\otimes\cdots\otimes\tau_{1},
\end{align}
where the catalyst $\tau^{CS}$ acts on composite subsystems $CC^{'}$ with
$C := A_{n-1} \ldots A_1$ and $C^{'} := B_{n-1} \ldots B_1$ ($A_i = A$ and $B_i = B$ for all $i$).
\end{lem}

Intuitively, this convex-split construction distributes $\rho$ uniformly across $n$ registers, ensuring that the catalyst experiences only negligible disturbance while enhancing the achievable fidelity of the distilled state.
In what follows, we introduce the embezzling-state–assisted (ESA) distillation protocol, which realizes catalytic enhancement through a distinct mechanism.

\begin{figure}[t]
    \centering
    \includegraphics[width=0.48\textwidth]{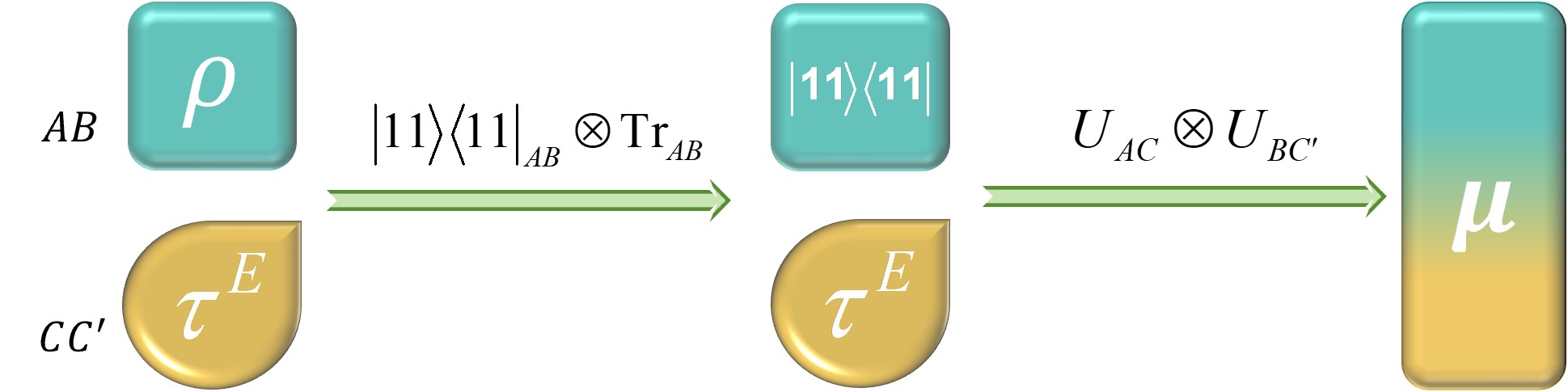}
    \caption{
    {\bf Embezzling-State–Assisted (ESA) Distillation.}
    The protocol begins by replacing the main system's state $\rho$ with $\ketbra{11}{11}$ and then applying the same unitary operation $U$ to the subsystems $AC$ and $BC'$, respectively. 
    The embezzling state $\tau^{E}$ (see Eq.~\eqref{eq:emb-state}) is prepared on the ancillary systems $C$ and $C'$. 
    The unitary $U$ is defined through its action on the computational basis as $U\ket{ij} = \ket{kl}$, where the indices satisfy $l = \lceil ((i-1)M + j)/d \rceil$ and $k = (i-1)M + j - (l-1)d$.
    }
    \label{fig:protocol-E}
\end{figure}
    
\begin{lem}
[{\bf ESA Distillation~\cite{Xing2024Teleportation}}]\label{lem:E}
For a bipartite state $\rho_{AB}$ of local dimension $d$ and a target threshold $\varepsilon>0$, the catalytic system $CC'$ is initialized in an embezzling state of the form
\begin{align}\label{eq:emb-state}
    \ket{\tau^{E}} 
    = 
    \frac{1}{\sqrt{c_M}}\sum_{j=1}^M\frac{1}{\sqrt{j}}\ket{jj},
\end{align}
where
\begin{align}
    c_M:= \sum_{j=1}^{M}\frac{1}{j},
\end{align}
and
\begin{align}\label{eq:M}
    M= 
    \left\lceil 
    d^{\frac{1}{1-\sqrt{1-\varepsilon}}} 
    \right\rceil.
\end{align}
This choice guarantees an entanglement fidelity
\begin{align}
    F
    \left(\Tr_{CC^{'}}[\Lambda^{E}(\rho\otimes \tau^{E})]\right) 
    \geqslant 
    1-\varepsilon,
\end{align}
with the corresponding LOCC operation $\Lambda^{E}$ illustrated in Fig.~\ref{fig:protocol-E}.
\end{lem}

The use of embezzling catalysts in entanglement distillation -- and, in particular, their application to quantum communication protocols such as teleportation -- has been demonstrated in recent work~\cite{Xing2024Teleportation}. 
These catalytic strategies transcend the limitations of conventional, catalyst-free distillation schemes, enabling near–unit-fidelity entanglement transformations even with finite resources. 
However, the embezzling catalyst itself inevitably undergoes a subtle yet cumulative degradation after each use, gradually diminishing its ability to facilitate further transformations. 
This observation raises a fundamental question central to the practicality of catalytic quantum communication: to what extent can a catalyst be reused before its advantage is lost? 
More precisely, how does this degradation constrain the number of effective reuses under realistic error tolerances? 
We address these questions in the following subsection.

\subsection{Reusability of Embezzling Catalysts}

In chemistry, the reuse of catalysts lies at the core of sustainable reaction design, enabling repeated enhancement of transformation efficiency without the continual expenditure of valuable resources.
Drawing a close parallel, we now turn to the quantum regime and examine the reusability of embezzling catalysts in entanglement distillation. 
These catalysts play an analogous role: they enable high-fidelity quantum transformations while tolerating small, controlled deviations from their initial state -- remaining useful as long as these deviations stay within an acceptable error range. 
Unlike their chemical counterparts, however, quantum catalysts are intrinsically fragile: each round of use introduces slight perturbations and exposure to environmental noise, leading to gradual degradation of their effectiveness. 
Quantifying this cumulative consumption is thus key to understanding the operational lifetime of embezzling catalysts and to establishing the fundamental limits on the sustainable use of auxiliary entangled resources.

\begin{figure*}
        \centering     
        \includegraphics[width=1\textwidth]{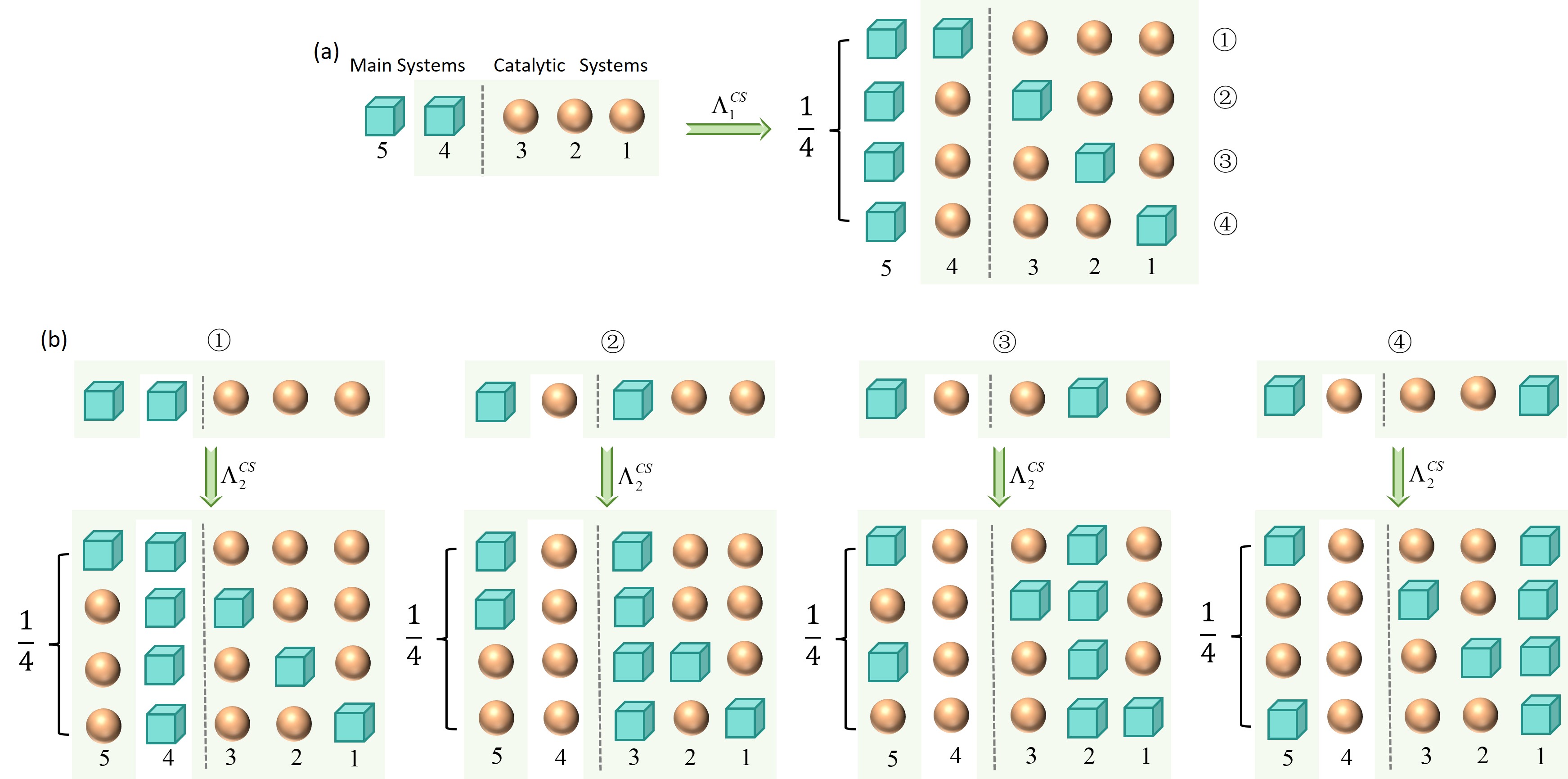}
        \caption{
        {\bf Convex-Split–Lemma–Assisted (CSLA) Distillation.}
        Representation of two rounds of the catalytic protocol $\Lambda^{CS}$ defined in Eq.~\eqref{eq:lmdCS} for the case $n=4$. 
        Indices $1$–$3$ label the catalytic subsystems, while $4$ and $5$ denote the main systems. 
        Green cubes represent the input states $\rho$, and yellow spheres indicate the components of the embezzling catalyst $\tau^{CS} = \tau_3 \otimes \tau_2 \otimes \tau_1$. 
        (a) In the first round, $\Lambda^{CS}_1$ acts on the main system $4$ and catalytic subsystems $1$–$3$, producing an equal mixture over four outcomes, with the reduced state $\rho^{CS,1}_4 = (\rho + 3\tau)/4$.
        (b) In the second round, $\Lambda^{CS}_2$ acts on the main system $5$ and the same catalytic subsystems. Each input branch generates four new outcomes, yielding a uniform mixture over $16$ states. The reduced state on system $5$ is $\rho^{CS,2}_5 = (7\rho + 9\tau)/16$, while system $4$ retains $\rho^{CS,2}_4 = [4(\rho + 3\tau)]/16=\rho^{CS,1}_4$, reflecting contributions carried over from the first round.
        }
        \label{fig:reuse-cs}
\end{figure*}

We now turn to the central theme of this work -- the reusability of embezzling catalysts in entanglement distillation. 
In each round of catalytic distillation, the target system achieves a higher entanglement fidelity, often approaching $1$, while the catalyst itself undergoes a minute but unavoidable perturbation.
Over successive uses, these perturbations accumulate, gradually eroding the catalyst's ability to sustain further transformations. 
It is therefore natural to define a finite operational lifetime for such a resource. 
Formally, we denote by $r$ the maximal number of rounds after which the catalytic distillation still produces an output state whose fidelity exceeds that of the original by a prescribed threshold $\epsilon$, namely,
\begin{align}\label{eq:distill-gain}    
    \Delta F = F(\rho^{\text{Cat}, t}) - F(\rho) > \epsilon,
\end{align}
where $F(\cdot)$ denotes the entanglement fidelity defined in Eq.~\eqref{eq:Entfidelity}, and $\rho^{\text{Cat}, t}$ represents the output state after the $t$-th catalytic distillation round. 
For all $t\leqslant r$, the above condition remains satisfied; 
once $t>r$, it is violated, marking the limit of the catalyst’s reusability.

Consider two types of embezzling catalytic protocols introduced in Lem.~\ref{lem:cs} (CSLA Distillation) and Lem.~\ref{lem:E} (ESA Distillation). 
Our analysis begins with the CSLA Distillation, which serves as a fundamental example for understanding catalytic behavior under repeated use. 
To systematically describe its evolution across multiple rounds, we introduce a consistent labeling scheme for the involved systems. 
The catalytic systems $CC'$ are partitioned into subsystems $A_iB_i$ for $i = 1, \dots, n-1$, while the system undergoing catalytic transformation in the $t$-th round, with $t \in [1, r]$, is labeled $A_{n+t-1}B_{n+t-1}$. 
An illustrative configuration with $n=4$ is shown in Fig.~\ref{fig:reuse-cs}. 
The following theorem quantifies the reusability limit of the convex-split catalyst, establishing the maximal number of rounds, $r_{CS}$, for which $\tau^{CS}$ continues to deliver a fidelity enhancement exceeding the threshold $\epsilon$ in every distillation cycle.

\begin{thm}
[{\bf Reusability of CSLA Catalyst}]
\label{thm:reuse-cs}
Let the initial noisy bipartite state be $\rho$ on systems $A$ and $B$, and fix a fidelity improvement threshold $\epsilon > 0$.
We employ an embezzling catalyst of the form $\tau^{CS} = \tau^{\otimes n-1}$ (see Lem.~\ref{lem:cs}).
After $r$ rounds of the catalytic distillation protocol, the resulting state on the system $n+r-1$ takes the form
\begin{align}\label{eq:r-state}
     \rho^{CS,r}_{n+r-1}=\frac{1}{n^r}\left((n^r-(n-1)^r)\rho + (n-1)^r\tau\right).
\end{align}  
Imposing the requirement that each round achieves a fidelity improvement $\Delta F > \epsilon$ (see Eq.~\eqref{eq:distill-gain}), the maximum number of reusable rounds, $r_{CS}$, is given by
\begin{align}\label{eq:drCS}
    r_{CS} =
    \left\lfloor 
    \frac{\log \epsilon - \log (F(\tau)-F(\rho)) }{\log(n-1)-\log n} 
    \right\rfloor.
\end{align}
This relation quantifies the operational lifetime of the CSLA catalyst $\tau^{CS}$ -- how long it can sustain a fidelity enhancement beyond the threshold $\epsilon$ before its catalytic advantage is fully depleted.
\end{thm}

Before proceeding with the proof, we first establish the notational conventions used throughout this section. 
In the CSLA distillation protocol (see Lem.~\ref{lem:cs}), the numbers of input states to be distilled sequentially -- namely, the number of copies of $\rho$ and the number of $\tau$ states composing the catalytic system $\tau^{CS}$ -- play a central role in analyzing the reusability of embezzling catalysts. 
To systematically label these systems, we adopt the following indexing convention. 
The catalytic subsystems are indexed from $1$ to $n-1$, and, following the layout in Fig.~\ref{fig:reuse-cs}(a), their subscripts are arranged from right to left. 
For instance, in the CSLA catalyst illustrated in Fig.~\ref{fig:reuse-cs}(a), the subsystem ordering is defined accordingly as
\begin{align}
    \tau^{CS}= \tau_3\otimes\tau_2\otimes\tau_1.
\end{align}
Meanwhile, the main system consists of two copies of the noisy entangled state $\rho$, denoted as $\rho^{\otimes 2}$. 
\begin{align}
    \rho^{\otimes2}=\rho_5\otimes\rho_4.
\end{align}
Hence, the initial configuration contains two $\rho$ states and three $\tau$ states, collectively represented as
\begin{align}\label{eq:index-ex1}
    \rho^{\otimes2}\otimes\tau^{CS}
    =
    \rho_5\otimes\rho_4\otimes
    \tau_3\otimes\tau_2\otimes\tau_1.
\end{align}
For brevity, we denote this configuration by the index pair $[2,3]$, i.e., 
\begin{align}
    \lambda
    \left(
    \rho_5\otimes\rho_4\otimes
    \tau_3\otimes\tau_2\otimes\tau_1
    \right)
    =
    [2,3],
\end{align}
where the first entry refers to the number of $\rho$ copies and the second to the number of $\tau$ components.

In the following discussion, we focus on the general case where the catalytic subsystems are indexed from $1$ to $n-1$, comprising a total of $n-1$ systems, while the main system is indexed from $n$ to $n+r-1$, comprising $r$ systems in total. 
For any composite state $\sigma$ acting jointly on both the main and catalytic systems, we denote its associated index pair by $\lambda(\sigma)$, and write it as
\begin{align}\label{eq:vr}
    \lambda(\sigma)=[x_1,\,x_2],
\end{align}
where the first entry $x_1$ represents the number of $\rho$ copies and the second entry $x_2$ denotes the number of $\tau$ components.
In deterministic scenarios -- such as the configuration shown in Eq.~\eqref{eq:index-ex1} -- the interpretation of $x_1$ and $x_2$ is straightforward.
However, in the probabilistic setting, the situation becomes more subtle: what does the pair $[x_1, x_2]$ signify when the system is described by a mixture of different configurations?
To clarify this, let us consider another representative example.
\begin{align}
    \frac{1}{2}\left(\rho_2\otimes\tau_1+\tau_2\otimes\rho_1\right).
\end{align}
In this case, the two components -- namely $\rho_2\otimes\tau_1$ and $\tau_2\otimes\rho_1$ -- are distinct, and thus we count them separately: the first corresponds to copy $\rho_2\otimes\tau_1$, and the second to copy $\tau_2\otimes\rho_1$. 
For simplicity, the normalization coefficient $1/2$ is omitted.
Thus, the index pair associated with this mixed configuration is therefore $[2,2]$, i.e.,
\begin{align}
    \lambda
    \left(
    \frac{1}{2}\left(\rho_2\otimes\tau_1+\tau_2\otimes\rho_1\right)
    \right)
    =[2,2].
\end{align}
According to our notational convention, whenever the components differ, each is counted individually; otherwise, they are counted only once.
For example, for the state $\rho = \rho/2 + \rho/2$, we still regard it as containing a single copy of $\rho$, and denote it by the index pair $[1,0]$.

Let us now apply this notational convention to streamline the analysis of the reusability of the CSLA catalyst.
For example, consider the configuration shown in Fig.~\ref{fig:reuse-cs}(a).
After the first round of the catalytic process -- namely, the implementation of $\Lambda^{CS}_1$ -- the resulting joint state of the global system, corresponding to the shaded region on the right-hand side of Fig.~\ref{fig:reuse-cs}(a), involves the main subsystem $4$ together with the catalytic subsystems $1$–$3$, and can be written as
\begin{align}
    \mu^1_{4C} =&
    \frac{1}{4} 
    (
    \rho_4\otimes\tau_3\otimes\tau_2\otimes\tau_1
    +
    \tau_4\otimes\rho_3\otimes\tau_2\otimes\tau_1
    \notag\\
    &+
    \tau_4\otimes\tau_3\otimes\rho_2\otimes\tau_1
    +
    \tau_4\otimes\tau_3\otimes\tau_2\otimes\rho_1
    ),
\end{align}
and its associated index pair is given by
\begin{align}
    \lambda(\mu^1_{4C})=[4,12].
\end{align}
Here, the global system spans subsystems $1$ to $4$ because, in the first round of catalytic entanglement distillation, only a single input state is involved.
Consequently, the $5$-th subsystem -- corresponding to the main system used in the second round -- is not yet included in the present consideration.

After completing the first round of catalytic entanglement distillation, we focus on the main subsystem $4$.
The resulting state can be expressed as
\begin{align}
    \rho^{CS,1}_4 =
    \frac{1}{4}(\rho_4 + 3\tau_4),
\end{align}
with the corresponding index pair written as
\begin{align}
    \lambda(\rho^{CS,1}_4) = [1,3].
\end{align}
Meanwhile, the catalytic system transforms into 
\begin{align}
    \tau^{CS,1}_C =
    \frac{1}{4} 
    (
    \tau_3\otimes\tau_2\otimes\tau_1
    +
    \rho_3\otimes\tau_2\otimes\tau_1
    &+
    \tau_3\otimes\rho_2\otimes\tau_1
    \notag\\
    &+
    \tau_3\otimes\tau_2\otimes\rho_1
    ),
\end{align}
whose index pair is characterized as
\begin{align}
    \lambda(\tau^{CS,1}_C) = [3,9].
\end{align}
These index pairs can also be directly verified from Fig.~\ref{fig:reuse-cs}(a).
Within the shaded region, the column to the left of the dashed line represents the main system, which contains one green cube and three yellow spheres, corresponding to the index pair $[1,3]$.
On the right-hand side of the dashed line, there are three columns comprising a total of three green cubes and nine yellow spheres.
In other words, the remaining catalytic system is characterized by the index pair $[3,9]$. 
If we further endow these index pairs with element-wise addition and subtraction, it is straightforward to check that they satisfy the following relation
\begin{align}
    \lambda(\tau^{CS,1}_C) 
    =\  
    \lambda(\mu^1_{4C}) - \lambda(\rho^{CS,1}_4).
\end{align}
In terms of indices, the relation takes the form
\begin{align}
    [3,9]= [4,12]- [1,3].
\end{align}
Having established the notational framework, we are now in a position to analyze the sequential reuse of the catalyst in entanglement distillation, thereby setting the stage for the proof of our Thm.~\ref{thm:reuse-cs}.

\begin{proof}
According to the established notational convention, the proof of Eq.~\eqref{eq:r-state} is equivalent to demonstrating the following relation, 
\begin{align}
    \lambda(\rho^{CS,r}_{n+r-1}) 
    = 
    \left[ n^r - (n-1)^r,\ (n-1)^r \right],
\end{align}
which we shall establish by mathematical induction.

For $r = 1$, after applying $\Lambda^{CS}_1$ once, the joint state of the main and catalytic systems is given by
\begin{align}
    \mu^1_{nC} 
    = 
    \frac{1}{n} \sum_{i=1}^n \tau_n \otimes \cdots \otimes \tau_{i+1} \otimes \rho_i \otimes \tau_{i-1} \otimes \cdots \otimes \tau_1.
\end{align}
Consequently, the reduced state on the main subsystem $n$ evolves into
\begin{align}
    \rho^{CS,1}_n = \frac{1}{n} \left( \rho + (n-1) \tau \right),
\end{align}
associated with the index pair
\begin{align}
    \lambda(\rho^{CS,1}_n) = [1, n-1].
\end{align}
For the inductive step, assume that for all $t<r$, the reduced state after $t$-th rounds satisfies
\begin{align}\label{eq:rho-main}
    \lambda(\rho^{CS,t}_{n+t-1}) 
    = 
    \left[ n^t - (n-1)^t,\ (n-1)^t \right].
\end{align}
We will show that the same form holds after the $r$-th round, thereby completing the inductive proof.

Furthermore, owing to the structure of the catalytic operation $\Lambda^{CS}_t$, which acts exclusively on subsystem $(n+t-1)$ together with the catalytic systems, the state of each fixed main subsystem $i$ evolves multiplicatively across successive rounds.
Formally,
\begin{align}\label{eq:rho-i}
    \lambda(\rho^{CS,t}_i) 
    = 
    n \cdot \lambda(\rho^{CS,t-1}_i), 
    \quad \forall i \in \{n, \ldots, n+r-2\}.
\end{align}
This multiplicative scaling originates from the recursive architecture of the protocol: 
each output state from the previous round branches into $n$ equivalent copies in the subsequent iteration (see Fig.~\ref{fig:reuse-cs}(b)).

After $(r-1)$-th rounds, the joint system evolves into a superposition of $n^{r-1}$ product states, denoted by $\mu^{r-1}_{(n+r-2)\cdots n C}$, each containing $r-1$ copies of $\rho$ and $n-1$ copies of $\tau$.
Accordingly, the associated index pair is
\begin{align}\label{eq:mu_r-1}
    \lambda(\mu^{r-1}_{(n+r-2)\cdots n C}) 
    = 
    n^{r-1} \cdot [r-1, n-1].
\end{align}
The index pair of the catalytic subsystem can then be represented as
\begin{align}
    \lambda(\tau^{CS,r-1}_C)
    = &\ 
    \lambda(\mu^{r-1}_{(n+r-2)\cdots n C}) 
    - 
    \sum_{t=1}^{r-1} \lambda(\rho^{CS,r-1}_{n+t-1})\\
    = &\
    (n-1) \cdot \left[ n^{r-1} - (n-1)^{r-1},\ (n-1)^{r-1} \right],       
\end{align}
where the second line follows directly from Eqs.~\eqref{eq:rho-main}, \eqref{eq:rho-i} and \eqref{eq:mu_r-1}.

In the $r$-th iteration, the main subsystem $(n+r-1)$ receives $n^{r-1}$ copies of $\rho$ together with the catalytic state from the preceding round, yielding
\begin{align}
    \lambda(\rho^{CS,r}_{n+r-1}) 
    =&\ 
    \lambda(\tau^{CS,r-1}_C) + [n^{r-1}, 0] \\
    =&\ 
    \left[ n^r - (n-1)^r,\ (n-1)^r \right],
\end{align}
which concludes the inductive step. Transforming this index pair representation back into the density operator form gives
\begin{align}\label{eq:CSLA-final}
    \rho^{CS,r}_{n+r-1} 
    = 
    \frac{1}{n^r} \left( (n^r - (n-1)^r) \rho + (n-1)^r \tau \right).
\end{align}
This result establishes the general expression for the output state on main system after $r$ rounds catalytic entanglement distillations.
         
\begin{figure*}[htp]
    \centering
    \includegraphics[width=1.0\textwidth]{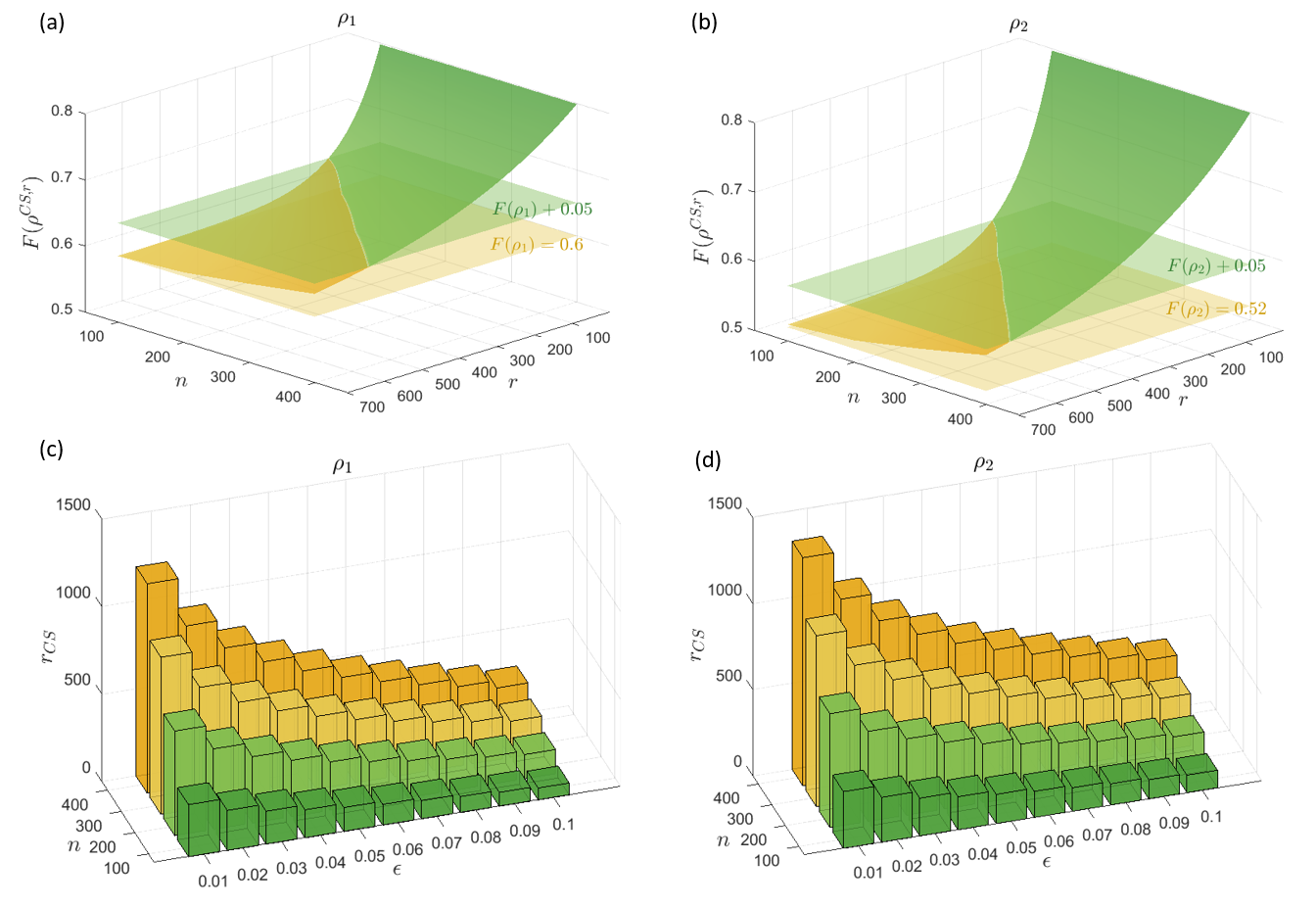}
    \caption{
    {\bf Reusability of CSLA Catalysts.}
    Numerical characterization of the catalytic performance and reusability of the CSLA catalyst $\tau^{CS}$ in entanglement distillation. Constructed through the convex-split-lemma as $\tau^{CS} = \tau^{\otimes (n-1)}$ (see Lem.~\ref{lem:cs}), this catalyst demonstrates sustained fidelity enhancement over successive rounds, revealing the operational lifetime of catalytic resources.
    Figures (a) and (b) show the fidelity $F(\rho^{CS,r})$ as a function of the distillation round $r$ and the parameter $n$ specifying the CSLA catalyst $\tau^{CS}= \tau^{\otimes (n-1)}$. 
    The yellow plane represents the fidelity of the original noisy state $\rho$, and the green plane indicates the performance threshold defining the boundary of effective catalyst reusability.
    Figures (c) and (d) depict the dependence of the maximum effective reuse rounds $r_{CS}$ (see Eq.~\eqref{eq:drCS}) on the fidelity improvement threshold $\epsilon$ and the catalyst size parameter $n$. 
    The observed scaling trend delineates the fundamental trade-off between enhanced distillation accuracy, increasing catalytic dimensionality, and the gradual loss of reusability inherent to the CSLA catalyst.
    }
    \label{fig:distillation-cs}
\end{figure*}
    
Having determined the explicit form of the main-system state after $r$ rounds of catalytic entanglement distillation, we next compute the resulting fidelity enhancement as
\begin{align}
    \Delta F 
    =
    F(\rho^{CS,r}_{n+r-1}) - F(\rho)
    =
    \left( \frac{n-1}{n} \right)^r \left( F(\tau) - F(\rho) \right).
\end{align}
To guarantee that $\Delta F > \epsilon$, the parameters must satisfy
\begin{align}
    \left( \frac{n-1}{n} \right)^r > \frac{\epsilon}{F(\tau) - F(\rho)}.
\end{align}
Taking logarithms on both sides and solving for $r$ gives
\begin{align}
    r < \frac{ \log \epsilon - \log \left( F(\tau) - F(\rho) \right) }{ \log(n-1) - \log n }.
\end{align}
Hence, the maximum integer number of permissible rounds $r_{CS}$ is
\begin{align}\label{eq:CSLA_max_rounds}
    r_{CS} = 
    \left\lfloor 
    \frac{ \log \epsilon - \log \left( F(\tau) - F(\rho) \right) }{ \log(n-1) - \log n } 
    \right\rfloor.
\end{align}
This concludes the proof.
     
\end{proof}

To investigate the reusability of catalytic resources under controlled conditions, we employ the theorem as a quantitative framework linking the fidelity improvement threshold $\epsilon$ and the catalytic system size $n$. 
Guided by this relation, we perform numerical simulations on two randomly selected states, $\rho_1$ and $\rho_2$, with initial entanglement fidelities of $0.6$ and $0.52$, respectively. 
The catalytic state $\tau^{CS}$ is assembled from a constituent state $\tau$ of fidelity $0.8$. 
We then examine the evolution of the output fidelity $F(\rho^{\text{Cat},r})$ over multiple catalytic distillation rounds $r$, and determine the maximum number of effective reuses $r_{CS}$ for a chosen $\epsilon$ and $n$ (see Eq.~\eqref{eq:drCS}), thereby quantifying how the fidelity improvement threshold and the catalyst size jointly constrain its operational lifetime.

As shown in Figs.~\ref{fig:distillation-cs}(a) and~\ref{fig:distillation-cs}(b), the output entanglement fidelity after catalytic distillation increases with the number of copies $n$, corresponding to a larger catalytic system, when the number of distillation rounds is held fixed. 
In contrast, for a fixed catalyst size $n$, the fidelity diminishes as the number of rounds increases, revealing the gradual accumulation of catalytic degradation and the fundamental trade-off between entanglement enhancement and reusability.
Extending this analysis, Figs.~\ref{fig:distillation-cs}(c) and~\ref{fig:distillation-cs}(d) characterize the reusability of the CSLA catalyst in terms of the fidelity improvement threshold $\epsilon$ and the catalyst size $n$. 
The results reveal that the maximum number of reusable rounds $r_{CS}$ increases with larger catalyst size and decreases with tighter fidelity requirements, capturing the quantitative structure of the fundamental trade-off governing catalytic entanglement distillation.

We now extend our analysis to the ESA Distillation (see Lem.~\ref{lem:E}). 
To consistently capture the system's evolution over multiple rounds, we introduce a unified labeling scheme for all subsystems involved. 
As depicted in Fig.~\ref{fig:Entdistillation}(c), the embezzling catalyst $\tau^{E}$ (see Eq.~\eqref{eq:emb-state}) is acting on registers $CC'$, while the main system subject to catalytic transformation in the $t$-th round ($t \in [1, r]$) is labeled $A_{t}B_{t}$. 
We denote by $\rho^{E,r}_{t}$ the state of the main system $A_{t}B_{t}$ after $r$ rounds of catalytic distillation. 
The theorem that follows provides the exact analytical form of the entanglement fidelity $F(\rho^{E,r}_r)$, establishing the quantitative basis for determining the catalyst's effective lifetime -- that is, the maximum number of rounds in which the embezzling catalyst remains operational.

\begin{thm}
[{\bf Reusability of ESA Catalyst}]
\label{thm:reuse-e}
Given a noisy bipartite state $\rho_{AB}$, we consider an embezzling catalyst of the form $\tau^{E} = 1/\sqrt{c_M} \sum_{j=1}^M 1/\sqrt{j} \ket{jj}$ (see Eq.~\eqref{eq:emb-state}). 
After $r$ rounds of catalytic distillation, the entanglement fidelity of the resulting state $\rho^{E,r}_r$ is given explicitly by
\begin{align}\label{eq:FUrhoEr}
    \ F(\rho^{E,r}_r)
    =
    &\frac{1}{d} + \frac{1}{dc_M} \sum^{\lceil \frac{M}{d^r}\rceil}_{s=1} \sum^{K^r_s}_{t=1}\sum^{K^r_{st}}_{h=1}
    x(s,t,h),
\end{align}
where the parameters are defined as
\begin{align}
    &x(s,t,h):=\frac{2}{\sqrt{(t+(s-1)d^r)(t+(s-1)d^r+hd^{r-1})}},\label{eq:para_1}\\
    &K^r_s:=\min\left\{d^r-d^{r-1},M-d^r, M-(s-1)d^r\right\},\label{eq:para_2}\\
    &K^r_{st}:=\min\left\{d-1,\left\lfloor \frac{d^r-t}{d^{r-1}}\right\rfloor, \left\lfloor \frac{M-(s-1)d^r-t}{d^{r-1}} \right\rfloor\right\}.\label{eq:para_3}
\end{align}  
\end{thm}

Each round of catalytic entanglement distillation involves four subsystems: $A_r$, $B_r$, $C$, and $C'$. 
To ensure a consistent description of the protocol, we introduce two notational conventions. The catalytic registers are collectively denoted as $\mathbf{C} := CC'$, representing the ancillary system that remains fixed throughout the process. 
The main system engaged in the $r$-th round is denoted as $\mathbf{r} := A_rB_r$, with $r$ labeling the round index. 
Under this convention, for example, $\ket{1}_{\mathbf{1}}$ corresponds to $\ket{11}_{A_1B_1}$, while $\ket{1}_{\mathbf{C}}$ denotes $\ket{11}_{CC'}$.
This notation provides a concise framework for expressing multi-round transformations and tracing the evolution of both the main and catalytic subsystems.

\begin{proof}
In multi-round catalytic entanglement distillation, the protocol $\Lambda^E$ (see Fig.~\ref{fig:protocol-E}) in the $t$-th round operates exclusively on the main system $\mathbf{t}$ and the catalytic registers $\mathbf{C}$, leaving all remaining subsystems unaffected. 
The protocol comprises two stages, the first of which is applied uniformly across all rounds: it re-initializes the main system $\mathbf{t}$ to the separable state $\ket{11}$ for each $t\in [1,r]$. 
After this initialization, the joint state of the composite system is given by
\begin{align}
   \frac{1}{\sqrt{c_M}}\sum^M_{j=1}
   \frac{\ket{1\cdots 1j}_{\bf r\cdots 1C}}{\sqrt{j}}.
\end{align}

We now analyze the second stage of the catalytic protocol $\Lambda^E$, proceeding round by round. 
In this step, the same unitary operation $U$ (see Fig.~\ref{fig:protocol-E}) is applied independently to the subsystems $AC$ and $BC'$. 
We then examine in detail how this operation drives the evolution of the joint quantum state across successive rounds.

The proof proceeds iteratively over the catalytic distillation rounds. 
In the first round, the unitary operation $U$ acts on the subsystem $\mathbf{1C}$, transforming its initial state into
\begin{align}
    \ket{\mu}_{\bf 1C} := U_{\bf 1C}\frac{1}{\sqrt{c_M}}\sum^M_{j=1}\frac{1}{\sqrt{j}}\ket{1j} \\
    = \frac{1}{\sqrt{c_M}}\sum^{M}_{j=1} \frac{1}{\sqrt{j}}\ket{j_1j^1_C}.
\end{align}
Here the transformed indices are defined by
\begin{align}
    j^1_C&:=\lceil \frac{j}{d} \rceil,\\
    j_1&:= j- (j^1_C-1)d = j- (\lceil \frac{j}{d} \rceil-1)d.
\end{align}
This expression explicitly characterizes how the unitary $U$ redistributes amplitude across the $d$-dimensional subspaces of the catalyst in the first catalytic distillation round.
In the second round, the unitary operation $U$ acts solely on subsystem $\mathbf{2C}$, leaving the state of $\mathbf{1}$ unchanged. 
The joint state of systems $\mathbf{21C}$ therefore evolves as
\begin{align}
    \ket{\mu}_{\bf 21C} :=&U_{\bf 2C}\frac{1}{\sqrt{c_M}}\sum^{M}_{j=1} \frac{1}{\sqrt{j}}\ket{1j_1j^1_C}  
    \\=&\frac{1}{\sqrt{c_M}}\sum^{M}_{j=1} \frac{1}{\sqrt{j}}\ket{j_2j_1j^2_C},
\end{align}
where
\begin{align}
    j^2_C&:=\lceil \frac{j^1_C}{d} \rceil=\lceil \frac{j}{d^2} \rceil,\\
    j_2&:= j^1_C- (j^2_C-1)d= \lceil \frac{j}{d} \rceil- (\lceil \frac{j}{d^2} \rceil-1)d.
\end{align}
Proceeding inductively, after $r$ rounds of the protocol, the joint state of the total system $\mathbf{r\cdots 1C}$ takes the form
\begin{align}
    \ket{\mu}_{\bf{r}\cdots\bf{1C}} 
    :=
    \frac{1}{\sqrt{c_M}}\sum^{M}_{j=1}
    \frac{1}{\sqrt{j}}\ket{j_r\cdots j_1j^r_C},
\end{align}
where the indices are defined recursively as
\begin{align}
    j^r_{C}&:=\left\lceil \frac{j}{d^r} \right\rceil,
    \\
    j_{s}&:=\left\lceil \frac{j}{d^{s-1}}\right\rceil - \left(\left\lceil \frac{j}{d^{s}}\right\rceil - 1\right) d,\ \forall s \in[1,r].\label{eq:jcrjs}
\end{align}
The reduced density matrix on the main system $\mathbf{r}$ is then obtained by tracing out all catalytic and preceding main subsystems, i.e., over $\mathbf{r-1}, \ldots, \mathbf{1}, \mathbf{C}$.
\begin{align}\label{eq:rhoE-ij}
    &\rho^{E,r}_{\mathbf{r}} \notag\\
    =&\frac{1}{c_M}\Tr_{\bf r-1\cdots 1C}\left[\sum^M_{i,j=1} \frac{1}{\sqrt{ij}}\ket{i_r\cdots i_1i^r_C}\bra{j_r\cdots j_1j^r_C}\right]\notag\\
    =&\frac{1}{c_M}\sum_{i,j} \frac{1}{\sqrt{ij}}
        \ketbra{x_{i}}{x_{j}}
\end{align}
with
\begin{align}\label{eq:x_ij}
    x_{m}:=
    \left\lceil \frac{m}{d^{r-1}}\right\rceil - \left(\left\lceil \frac{m}{d^{r}}\right\rceil - 1\right) d,\quad m\in \{i,j\}.  
\end{align}
Note that our notational convention is applied here; thus, the state labeled by $\mathbf{r}$ acts jointly on subsystems $A_r$ and $B_r$. 
Written in full, the state $\ket{x_{i}}$ reads
\begin{align}
    \ket{x_{i}}_{\mathbf{r}}=\ket{x_{i},x_{i}}_{A_r B_r}.
\end{align}
The indices $i$ and $j$ appearing Eq.~\eqref{eq:rhoE-ij} are constrained by
\begin{align}
    \left\lceil \frac{i}{d^r} \right\rceil &= \left\lceil \frac{j}{d^r} \right\rceil, \label{eq:icjc} \\
     \left\lceil \frac{i}{d^{s-1}}\right\rceil - \left(\left\lceil \frac{i}{d^{s}}\right\rceil - 1\right) d 
   &=
    \left\lceil \frac{j}{d^{s-1}}\right\rceil - \left(\left\lceil \frac{j}{d^{s}}\right\rceil - 1\right) d, \label{eq:isjs}
\end{align}
where $s \in[1,r-1]$.
From the first constraint in Eq.~\eqref{eq:icjc}, the indices $i$ and $j$ used in Eq.~\eqref{eq:rhoE-ij} must satisfy
\begin{align}\label{eq:condition1}
    |i-j|<d^r,
\end{align} 
which follows from the stepwise nature of the ceiling function.
Specifically, $\left\lceil i/d^r \right\rceil$ increments by one each time $i$ increases by $d^r$, partitioning the integer sequence into contiguous blocks of width $d^r$. 
Furthermore, the function $j_s$ (see Eq.~\eqref{eq:jcrjs}) is periodic with period $d^s$ for each $s \in [1, r-1]$.
Since the largest common period among these $r-1$ functions is $d^{r-1}$, any pair of indices $i$ and $j$ satisfying Eq.~\eqref{eq:isjs} must differ by an integer multiple of $d^{r-1}$, namely,
\begin{align}\label{eq:condition2}
    |i-j| = hd^{r-1}, \quad  h\in \mathbb{N}.
\end{align}
It then follows directly that the entanglement fidelity of $\rho^{E,r}_{\bf r}$ is given by Eq.~\eqref{eq:FUrhoEr}, with the relevant parameters specified in Eqs.~\eqref{eq:para_1}-\eqref{eq:para_3}, thereby completing the proof.

\end{proof}

\begin{figure}[t]
    \centering
    \includegraphics[width=0.48\textwidth]{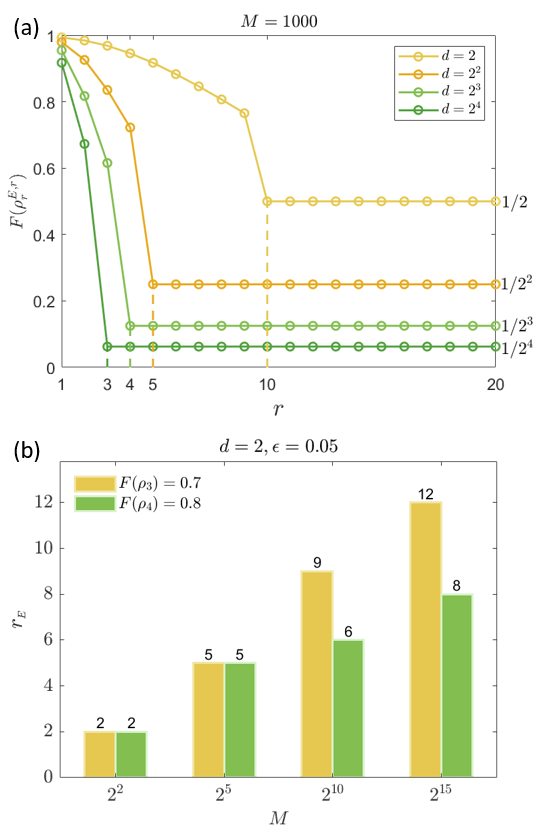}
    \caption{
    {\bf Reusability of ESA Catalysts.}
    The catalyst $\tau^{E}$ is constructed from an embezzling state with Schmidt rank $M$ (see Lem.~\ref{lem:E}).
    (a) Entanglement fidelity $F(\rho^{E,r}_r)$ as a function of the number of catalytic entanglement distillation rounds $r$ for different target dimensions $d$. 
    The fidelity exhibits a gradual decay with repeated use, converging to the limit $1/d$ when the number of rounds exceeds $\lceil\log_d M\rceil$.
    (b) Dependence of the maximum effective reuse rounds $r_E$ on the catalyst's Schmidt rank $M$, evaluated for $d=2$ and a fidelity-gain threshold $\epsilon=0.05$. 
    The scaling highlights the extended operational lifetime afforded by catalysts with more entanglement.
    }
    \label{fig:reuse_E}
\end{figure}   

\begin{figure*}
    \centering
    \includegraphics[width=0.98\textwidth]{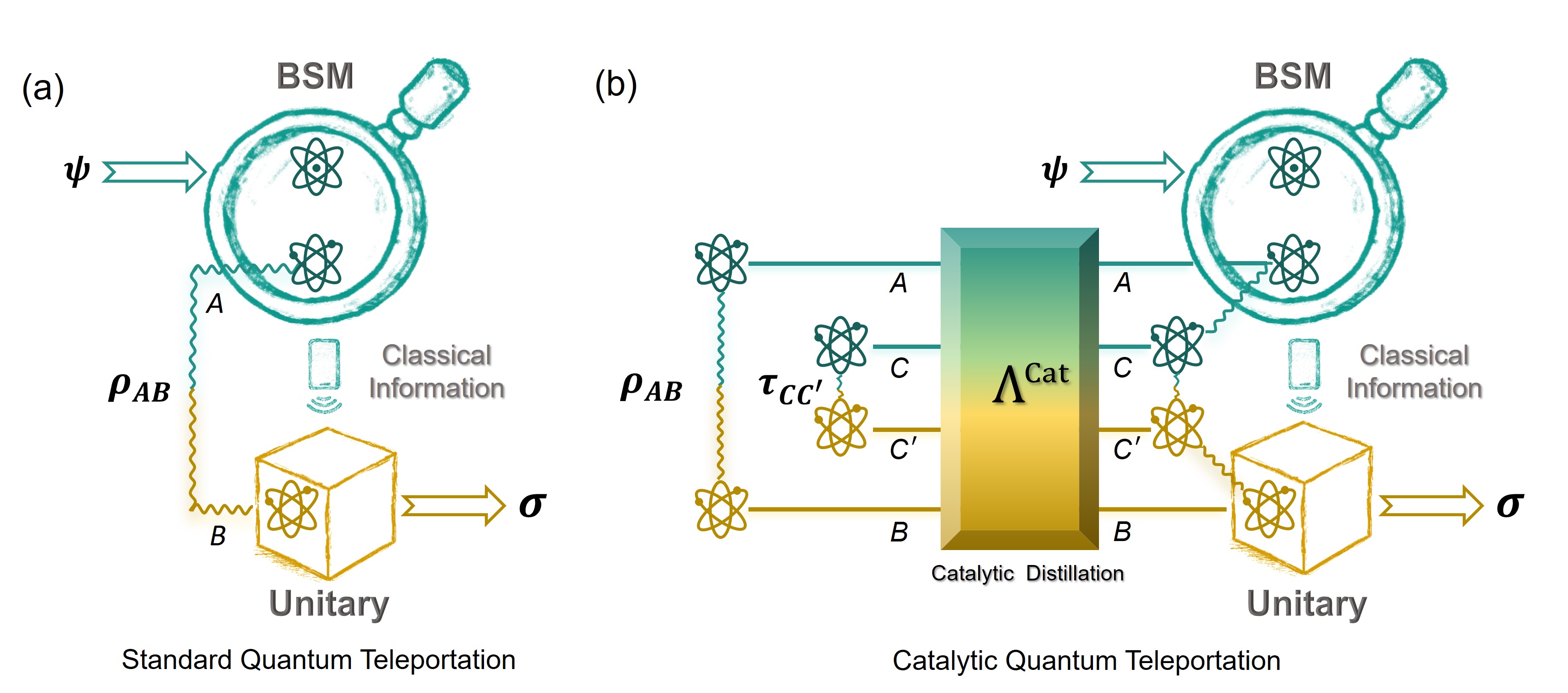}
    \caption{{\bf Catalytic Quantum Teleportation.} 
    (a) Standard teleportation of an unknown quantum state $\psi$ from sender (Alice or simply A) to receiver (Bob or simply B) using a pre-shared entangled state $\rho_{AB}$. 
    The process is implemented via a local operations and classical communication (LOCC) protocol $\Theta_0$, consisting of a Bell measurement (BSM) on Alice's side followed by a conditional unitary operation on Bob's side.
    (b) Catalytic teleportation assisted by an embezzling catalyst $\tau_{CC'}$. 
    The catalytic map $\Lambda^{\text{Cat}}$, introduced in Sec.~\ref{sec:distillation}, enhances the entanglement fidelity of the shared state -- and hence the teleportation fidelity -- while preserving the catalyst for subsequent reuse.
    }
    \label{fig:teleportation}
\end{figure*}

This theorem provides a full characterization of how the main system's quantum state $\rho^{E,r}_{\mathbf{r}}$ (see Eq.~\eqref{eq:rhoE-ij}) evolves through successive rounds of catalytic distillation. 
The corresponding numerical results are shown in Fig.~\ref{fig:reuse_E}. As illustrated in Fig.~\ref{fig:reuse_E}(a), for a fixed Schmidt rank $M=10^3$ of the ESA catalyst $\tau^E$ (see Eq.~\eqref{eq:emb-state}), the entanglement fidelity of the output state $\rho^{E,r}_{\mathbf{r}}$ decreases monotonically with the number of rounds $r$, eventually converging to the baseline value $1/d$ once $r \geqslant \lceil \log_d M \rceil$. 
This scaling behavior reflects the finite operational lifetime of the catalyst: beyond a limited number of uses, its ability to sustain fidelity enhancement above a prescribed threshold $\epsilon$ is inevitably lost.

To further investigate the relationship between the dimension of the ESA catalyst $\tau^E$ (see Eq.~\eqref{eq:emb-state}) and its reusability, we consider two randomly generated input states, $\rho_3$ and $\rho_4$, with initial entanglement fidelities of $0.7$ and $0.8$, respectively (the subscripts are used to distinguish these states from those analyzed for the CSLA catalyst). 
As illustrated in Fig.~\ref{fig:reuse_E}(b), for a fixed fidelity improvement threshold $\epsilon=0.05$, increasing the dimension $M$ of the ESA catalyst -- equivalently, its intrinsic entanglement -- extends the number of distillation rounds for which the output fidelity remains above the target threshold $\epsilon$, consistent with physical intuition. 
However, the quantitative results show that even gaining just a few additional usable rounds requires an enormous increase in catalyst dimension and entanglement. 
This steep scaling underscores the intrinsic limitation of ESA catalysts and motivates the development of more reusable and resource-efficient catalytic architectures, which we leave for future exploration. 

It is worth noting that, unlike in the analysis of the CSLA catalyst $\tau^{CS}$ (see Eq.~\eqref{eq:CSLA-Catalyst}), where both the exact form of the target state $\rho^{CS,r}_{n+r-1}$ (see Eq.~\eqref{eq:CSLA-final}) after $r$ rounds of catalytic entanglement distillation and the closed-form expression for the maximal number of effective rounds $r_{CS}$ (see Eq.~\eqref{eq:CSLA_max_rounds}) were derived, the case of the ESA catalyst $\tau^E$ (see Eq.~\eqref{eq:emb-state}) is more intricate. 
Here, we are able to obtain only the exact expression for the main system's state $\rho^{E,r}_{\mathbf{r}}$ (see Eq.~\eqref{eq:rhoE-ij}) after $r$ rounds of catalytic distillation. 
This limitation arises from the mathematically complex structure of the ESA distillation protocol. Nevertheless, this does not prevent us from determining the maximal number of rounds for which the entanglement fidelity remains above the prescribed threshold, as illustrated in Fig.~\ref{fig:reuse_E}(b).

\section{Catalytic Teleportation}
\label{sec:teleportation}

Having established the quantitative framework for catalyst reusability in entanglement distillation, we now turn to one of its most revealing operational consequences -- catalytic quantum teleportation~\cite{PhysRevLett.127.080502,Xing2024Teleportation} (see Fig.~\ref{fig:teleportation}). 
Quantum teleportation relies on shared entanglement as a consumable resource to faithfully transfer an unknown quantum state between distant parties. 
Ideally, each round of teleportation assumes access to a perfectly maximally entangled state between the sender and receiver. 
In practice, however, environmental noise inevitably degrades this shared resource, reducing the teleportation fidelity. 
Conventional approaches mitigate such degradation through entanglement distillation or purification, but their performance is fundamentally constrained when the total amount of entanglement is fixed. 

Recent advances have shown that quantum catalysts -- including correlated catalysts such as Duan's state~\cite{PhysRevLett.127.080502} and embezzling catalysts such as the CSLA and ESA catalysts~\cite{Xing2024Teleportation} -- can boost teleportation fidelity beyond conventional limits. 
Yet, the extent to which these catalytic resources can be reused across multiple rounds of teleportation has remained largely unexplored. 
Here, we address this question by analyzing the reusability of embezzling catalysts in quantum teleportation, establishing when and how they continue to confer an advantage over standard schemes. 
This extension unifies the concepts of catalytic entanglement distillation and catalytic quantum communication, uncovering a new operational role of quantum catalysts as enablers of sustained, high-fidelity state transfer across successive uses. 
The following analysis incorporates the previously derived reusability bounds into the teleportation framework, quantifying the precise regimes in which catalytic teleportation maintains its superiority.

Standard quantum teleportation (see Fig.~\ref{fig:teleportation}(a)) employs a pre-shared entangled state $\rho_{AB}$ to allow the sender (Alice) to transmit an unknown quantum state to the receiver (Bob). 
In the protocol, Alice performs a Bell measurement on the input (message) state $|\psi\rangle$ and her half of the entangled pair, then communicates the outcome to Bob through a classical channel. 
Upon receiving this information, Bob applies the corresponding unitary correction to his subsystem, thereby reconstructing the original state. 
The overall performance of teleportation is quantified by the average fidelity~\cite{PhysRevA.60.1888,xing2023fundamental},
\begin{align}
    f(\rho_{AB}):=\int d\psi\left\langle\psi\right|\Theta_0(\psi_R\otimes\rho_{AB})|\psi\rangle, 
\end{align}
where $\Theta_0$ denotes the standard teleportation protocol, encompassing both the Bell measurement and the conditional unitary operations. 
Crucially, this fidelity depends solely on the entanglement fidelity of the shared state $\rho_{AB}$, given by
\begin{align}\label{eq:tel-fidelity}
    f(\rho_{AB})=\frac{F(\rho_{AB})d+1}{d+1},
\end{align}
where $d$ is the local Hilbert space dimension of systems $A$ and $B$, and $F$ denotes the entanglement fidelity defined in Eq.~\eqref{eq:Entfidelity}. 
This relationship directly links the quality of teleportation to the degree of entanglement retained in the shared resource, forming the foundation for analyzing how catalytic processes can sustain or enhance teleportation fidelity across successive uses.

To enhance the performance of teleportation, Ref.~\cite{Xing2024Teleportation} introduced the concept of teleportation with embezzling catalyst, wherein an embezzling catalyst $\tau_{CC'}$ assists the protocol (see Fig.~\ref{fig:teleportation}(b)). 
The key idea is that the catalyst enables an enhancement of the single-shot entanglement distillation, thereby improving the fidelity of the shared entangled resource used for teleportation. 
The corresponding catalytic teleportation fidelity is given by
\begin{align}
    f_c(\rho_{AB})
    := &
    \int d\psi\left\langle\psi|\Theta(\psi_R\otimes\rho_{AB}\otimes\tau_{CC'})|\psi\right\rangle\\
    = & \frac{F(\rho^{\text{Cat}}_{AB})d+1}{d+1},
\end{align}
where $\Theta$ denotes the composite operation consisting of the catalytic distillation map $\Lambda^{\text{Cat}}$ followed by the standard teleportation protocol $\Theta_0$. 
The state $\rho^{\text{Cat}}_{AB}$ represents the effectively distilled shared state with enhanced entanglement fidelity, achieved through the use of the embezzling catalyst $\tau_{CC'}$.

The enhancement of teleportation enabled by embezzling catalysts has been detailed in Ref.~\cite{Xing2024Teleportation}; however, the central feature of catalytic processes -- their reusability -- has so far remained unexplored. 
We now extend the framework of catalyst reusability, established for entanglement distillation in this work, to the quantum teleportation scenario, thereby filling this conceptual and operational gap. 
This extension is made possible by a key correspondence: the average teleportation fidelity is uniquely determined by the entanglement fidelity of the shared bipartite state (see Eq.~\eqref{eq:tel-fidelity}). 
Leveraging this relationship, we formulate a criterion for the effective reuse of catalysts in multi-round teleportation, requiring that the average fidelity of the shared state in the $r$-th round satisfies
\begin{align}\label{eq:tel-deltaf}
    \Delta f = f(\rho^{\text{Cat},r}) - f(\rho) > \epsilon,
\end{align}
where $\epsilon$ denotes a prescribed fidelity improvement threshold, $\rho^{\text{Cat},r}$ is the shared state after the $r$-th catalytic teleportation, and $\rho$ is the initial shared entangled state.

Within this framework, Thms.~\ref{thm:reuse-cs} and~\ref{thm:reuse-e} provide the foundation for determining the maximum number of reusable rounds in catalytic teleportation. 
Specifically, we derive upper bounds for two distinct classes of catalysts -- those assisted by the convex-split lemma (CSLA catalysts) and those based on embezzling states (ESA catalysts). 
The corresponding results are summarized in the following two corollaries. 
We begin by examining the case of quantum teleportation assisted by a CSLA catalyst and analyze its reusability.
 
\begin{figure*}
    \centering
    \includegraphics[width=1.0\textwidth]{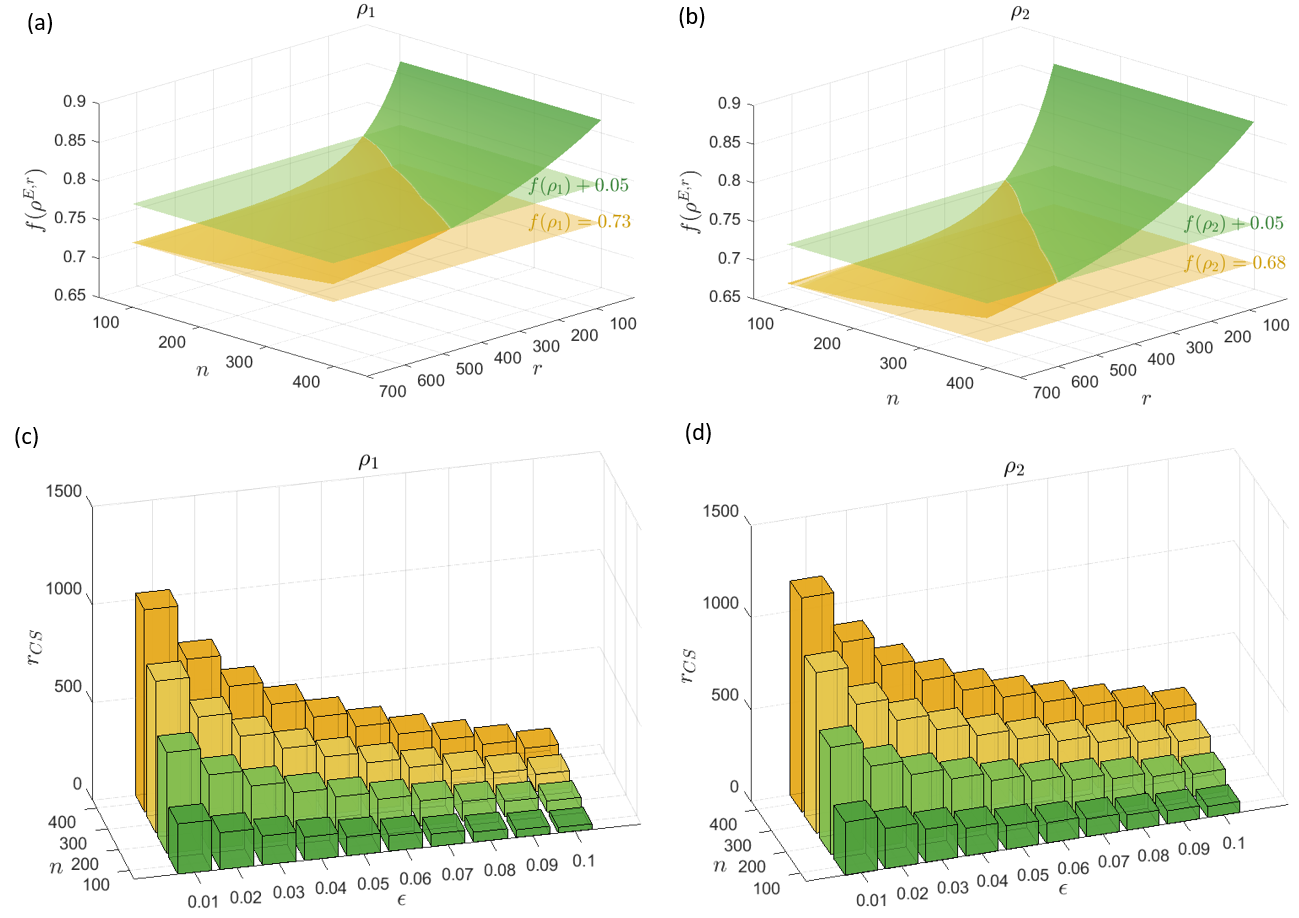}
    \caption{{\bf Reusability of CSLA Catalysts in Teleportation.}
    The catalyst is constructed via the convex-split lemma as $\tau^{CS} := \tau^{\otimes n-1}$ (see Lem.~\ref{lem:cs}).   Figures (a) and (b) display the average fidelity $f(\rho^{CS,r})$ as a function of both the catalyst dimension parameter $n$ and the teleportation round $r$. The yellow plane indicates the average fidelity for the initial noisy state $\rho$, while the green plane marks the performance threshold for effective catalyst reuse. Figures (c) and (d) illustrate how the maximum  effective reuse rounds $r_{CS}$ (see Eq.~\eqref{eq:trCS}) varies with the fidelity improvement threshold $\epsilon$ and the catalyst size parameter $n$. These scaling trends reveal a three-way trade-off among teleportation accuracy, catalyst dimensionality, and reusable lifetime inherent to the CSLA catalyst. The states $\rho_1$ and $\rho_2$ are identical to those introduced in Fig.~\ref{fig:distillation-cs}.  
    }
    \label{fig:teleportation_CS}
\end{figure*}

\begin{cor}
[{\bf CSLA Teleportation}]
\label{cor:reuse-cs}
For teleportation with an initial state $\rho$, we introduce a fidelity improvement threshold $\epsilon > 0$ to quantify the operational gain per round. 
Considering the CSLA catalyst $\tau^{CS}$ (see Lem.~\ref{lem:cs}), the maximum number of rounds over which $\tau^{CS}$ can be effectively reused while maintaining an average fidelity enhancement $\Delta f > \epsilon$ (see Eq.~\eqref{eq:tel-deltaf}) is given by
\begin{align}\label{eq:trCS}
    r_{CS}=\left\lfloor \frac{\log \frac{(d+1)\epsilon}{d} - \log (F(\tau)-F(\rho)) }{\log(n-1)-\log n} \right\rfloor.
\end{align}
\end{cor}

Building on the above corollary, we now investigate the operational reusability of embezzling catalysts in quantum teleportation -- that is, how many rounds of teleportation a given catalyst can sustain, or equivalently, how large a catalyst is required to support a prescribed number of rounds. 
For consistency, we employ the same entangled states as those analyzed in Fig.~\ref{fig:distillation-cs}, which serve as the shared resource between the sender and receiver. 
Our analysis proceeds from two complementary perspectives. 
First, we fix the fidelity improvement threshold at $\epsilon = 0.05$ and evaluate the average teleportation fidelity as a function of the catalyst size $n$ and the number of teleportation rounds $r$. 
The corresponding numerical results, shown in Figs.~\ref{fig:teleportation_CS}(a) and~\ref{fig:teleportation_CS}(b), reveal two clear trends: as the number of rounds increases, the average fidelity gradually decreases due to cumulative catalytic degradation; 
conversely, increasing the catalyst size $n$ leads to a higher achievable fidelity, reflecting the enhanced catalytic capacity.
Second, we investigate how the fidelity improvement threshold $\epsilon$ and catalyst size $n$ jointly influence the maximum number of catalytic teleportation rounds that maintain an improvement above the prescribed threshold. 
As illustrated in Figs.~\ref{fig:teleportation_CS}(c) and~\ref{fig:teleportation_CS}(d), a smaller threshold $\epsilon$ allows for a greater number of reusable rounds, while larger catalysts exhibit longer operational lifetimes, confirming the scaling behavior predicted by our analytical bounds.

\begin{cor}
[{\bf ESA Teleportation}]
\label{cor:reuse-e}
Consider a catalytic quantum teleportation protocol using a shared bipartite state $\rho$ and the ESA catalyst $\tau^{E}$ (see Lem.~\ref{lem:E}). 
For a prescribed fidelity improvement threshold $\epsilon > 0$, the protocol is required to satisfy $\Delta f>\epsilon$ (see Eq.~\eqref{eq:tel-deltaf}). 
Under this condition, the embezzling catalyst $\tau^{E}$ can be effectively reused for at most $r_E$ rounds, where the number of permissible reuse rounds is bounded by
\begin{align}\label{eq:trE}
    r_E=\max\left\{r: 
    \frac{1}{d} + \frac{1}{dc_M} \sum^{\lceil \frac{M}{d^r}\rceil}_{s=1} \sum^{K^r_s}_{t=1}\sum^{K^r_{st}}_{h=1}
    x(s,t,h) - 
    F(\rho)\right.\notag\\
    > \left.\frac{(d+1)\epsilon}{d} \right\},
\end{align}   
where all parameters, i.e., $x(s,t,h)$, $K^r_s$, and $K^r_{st}$, are defined in Eqs.~\eqref{eq:para_1}–\eqref{eq:para_3}.
\end{cor}

\begin{figure}
    \centering
    \includegraphics[width=0.48\textwidth]{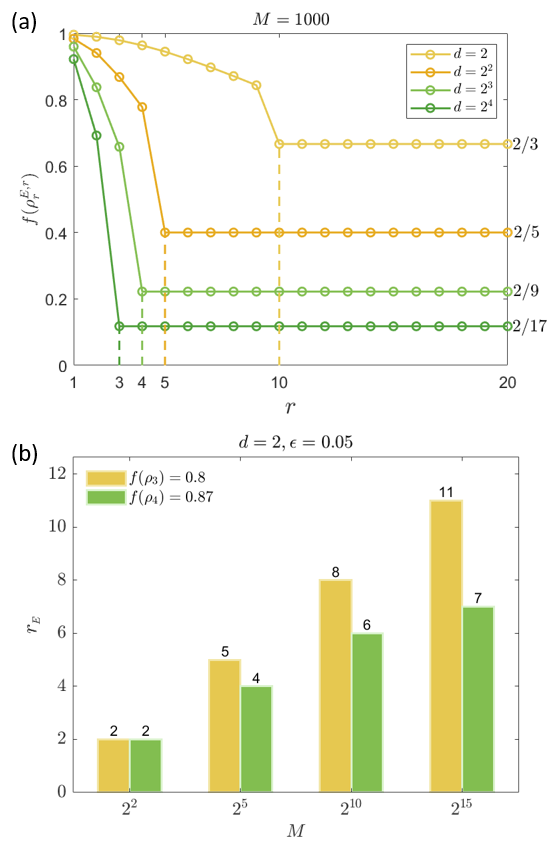}
    \caption{{\bf Reusability of ESA Catalysts in Teleportation.}
    The catalyst $\tau^{E}$ is constructed from an embezzling state with Schmidt rank $M$ (see Lem.~\ref{lem:E}).
    (a) The average fidelity $f(\rho^{E,r}_r)$ is plotted as a function of the catalytic teleportation round $r$ for various target dimensions $d$. 
    The average fidelity of catalytic teleportation decreases gradually with successive reuse of the catalyst and asymptotically approaches the limiting value $2/(d+1)$ once $r \geqslant \lceil \log_d M \rceil$.
    (b) The maximum number of effective reuse rounds $r_E$ is shown as a function of the catalyst's Schmidt rank $M$, for $d=2$ and a fidelity improvement threshold $\epsilon = 0.05$. 
    The scaling trend reveals that catalysts with larger Schmidt rank, or equivalently higher entanglement, sustain longer operational lifetimes. The states $\rho_3$ and $\rho_4$ are identical to those introduced in in Fig.~\ref{fig:reuse_E}.
    }
    \label{fig:teleportation_E}
\end{figure}    

Equipped with the results described in Cor.~\ref{cor:reuse-e}, we perform numerical simulations of catalytic quantum teleportation, in which the same catalyst is reused across multiple rounds. 
The corresponding results are summarized in Fig.~\ref{fig:teleportation_E}. 
In Fig.~\ref{fig:teleportation_E}(a), we fix the Schmidt rank of the embezzling state $\tau^{E}$ to $M=1000$ and explore how the average teleportation fidelity varies with the number of catalytic rounds $r$. 
As $r$ increases, the fidelity exhibits a gradual decline, eventually converging to the classical communication limit.
Fig.~\ref{fig:teleportation_E}(b) shows the results for the same input states as in Fig.~\ref{fig:reuse_E}. 
For fixed target dimension $d$ and fidelity gain threshold $\epsilon$, the catalyst's reusability -- quantified by the maximum number of rounds for which catalytic teleportation continues to outperform conventional teleportation above the target
threshold -- is determined by the Schmidt rank of the embezzling state.


\section{Discussions}
\label{sec:conclusion}

Quantum catalysts offer a fundamentally new avenue for enhancing the performance of quantum information processing. 
Depending on their operational setup, catalysts can be broadly classified into distinct categories. 
Exact catalysts improve performance while remaining completely separable from the main system and perfectly restored after the catalytic process. 
However, for a given quantum task, there is generally no systematic way to determine whether a useful catalyst exists or how to construct one that yields a measurable advantage. 
In contrast, embezzling catalysts can elevate performance to near-perfect levels, but at the cost of a slight disturbance to their own state. 
This gradual degradation ultimately limits the number of times such catalysts can be reused before losing their catalytic power.

In this work, we focus on a central primitive in quantum communication -- entanglement distillation -- to explore the reusability of quantum catalysts. 
We systematically analyze how key parameters, including the fidelity improvement threshold and the catalyst size, influence the operational lifetime of catalytic advantage. 
Our results establish an explicit upper bound on the maximal number of catalytic rounds for which performance remains superior to that of the non-catalytic scenario, given a prescribed improvement threshold.
As a direct application, we further extend these findings to catalytic quantum teleportation, highlighting the broader implications of catalytic reusability for quantum communication protocols.

Looking forward, our framework for quantifying catalytic reusability lays the groundwork for a broader theory of quantum resources -- one that treats catalysts not as ideal auxiliaries, but as evolving agents subject to degradation, recovery, and adaptation. 
Extending this perspective to multipartite networks~\cite{PhysRevLett.125.240505,PhysRevLett.132.080201,PhysRevLett.134.080202}, continuous-variable architectures~\cite{PhysRevX.14.011013,Fadel_2025,dy4m-gq5c}, and non-Markovian environments~\cite{PhysRevResearch.3.023077,PhysRevLett.130.240201,liu2024nonmarkoviannoisesuppressionsimplified} may reveal new pathways for sustaining catalytic functionality. 
The integration of data-driven optimization could further enable the design of catalysts tailored to specific quantum information processing or communication tasks. 
Yet, the operational mechanisms and ultimate performance limits of catalytic assistance in practical applications -- ranging from quantum key distribution~\cite{RevModPhys.92.025002} and repeaters~\cite{PhysRevLett.81.5932} to quantum error correction~\cite{6778074,fang2024surpassingfundamentallimitsdistillation} -- remain largely uncharted, delineating compelling frontiers for future exploration.

\section*{Acknowledgments}
This research is supported by the Stable Supporting Fund of Acoustic Science and Technology Laboratory (No. JCKYS2024604SSJS001) and the Fundamental Research Funds for the Central Universities (Nos. 3072024XX2401, 3072025CFJ2407, 3072025YC2401 and 3072025YC2402). 
Bing Yu is funded by the Startup Funding of Guangdong Polytechnic Normal University (No. 2021SDKYA178) and Key Laboratory of Computational Science and the Application of Hainan Province (No. JSKX202503).
Yunlong Xiao acknowledges support from A*STAR under its Career Development Fund (C243512002).



\bibliography{Bib}
\end{document}